\documentclass[reprint,nofootinbib]{revtex4-1}
\usepackage{amsmath}
\usepackage{braket}
\usepackage{graphicx}
\usepackage{color}
\begin{document}

\title{Dynamical Mean-Field Theory Simulations with the Adaptive~Sampling~Configuration~Interaction Method}
\date{\today}
\author{Carlos Mejuto-Zaera$^{\dagger}$, Norm M. Tubman$^{*}$, K. Birgitta Whaley\\
$^{\dagger}$ carlos$\_$mejutozaera@berkeley.edu, $^{*}$ norm.m.tubman@gmail.com}
\affiliation{University of California, Berkeley, California 94720, United States}
\begin{abstract}
In the pursuit of accurate descriptions of strongly correlated quantum many-body systems, dynamical mean-field theory (DMFT) has been an invaluable tool for elucidating the spectral properties and quantum phases of both phenomenological models and \emph{ab initio} descriptions of real materials. Key to the DMFT process is the self-consistent map of the original system into an Anderson impurity model, the ground state of which is computed using an impurity solver. The power of the  method is thus limited by the complexity of the impurity model the solver can handle. Simulating realistic systems generally requires many correlated sites. 
By adapting the recently proposed adaptive sampling configuration interaction (ASCI) method as an impurity solver, we enable much more efficient zero temperature DMFT simulations. The key feature of the ASCI method is that it selects only the most relevant Hilbert space degrees of freedom to describe the ground state. This reduces the numerical complexity of the calculation, which will allow us to pursue future DMFT simulations with more correlated impurity sites than in previous works. Here we present the ASCI-DMFT method and example calculations on the one-dimensional and two-dimensional Hubbard models that exemplify its efficient convergence and timing properties. We show that the ASCI approach is several orders of magnitude faster than the current best published ground state DMFT simulations, which allows us to study the bath discretization error in simulations with small clusters, as well as to address cluster sizes beyond the current state of the art.  Our approach can also be adapted for other embedding methods such as density matrix embedding theory and self-energy embedding theory.
\end{abstract}

\maketitle

\section{Introduction}

The systematic study of the properties of strongly correlated many-electron systems remains one of the main areas of research in condensed matter physics. In this regard, dynamical mean-field theory (DMFT) has been successfully applied to study metal-insulator phase transitions~\cite{Capone2004, Park2008}, exotic quantum phases of matter~\cite{He2012, Simons2015}, critical exponents in quantum field theories \cite{Akerlund2013}, the volume expansion in Plutonium \cite{Kotliar2001b} and high temperature superconductivity \cite{Lichtenstein2000,Go2015} among other. DMFT works by mapping the many body system of interest self-consistently into an  Anderson impurity Hamiltonian \cite{Kotliar1996}.
A central feature of the method is the impurity solver, which finds the ground state of the impurity model. Numerically exact solvers based on Monte Carlo approaches \cite{Werner2006, Haule2007, Gull2011} can be used in the finite temperature case, but these have difficulties converging to the $T = 0$ regime.  

One of the greatest challenges to expand the applicability of zero temperature DMFT is to find an efficient representation of the ground state wave function of the impurity system~\cite{Caffarel1994,Bulla1999,Garcia2004, Nishimoto2004, Ganahl2015, Wolf2015,Zgid2011, Zgid2012, Go2017,Lu2014}. 
 Simulations with configuration interaction (CI)~\cite{Szabo2012} based approaches have been used in zero temperature studies.  These approaches attempt to identify a  subspace in which to find the ground state and have had some success in treating strongly correlated systems~\cite{Bender1969, Bryant1987, Boustani1987, Bonacic1993}.  One of these methods, CI singles and doubles (CISD), has been used in an attempt to increase the size of the systems that can be simulated~\cite{Zgid2011, Zgid2012}. 
Additionally an iterative CISD has been considered as a DMFT solver~\cite{Lu2014,Go2017}, which has also been successful, but somewhat computationally expensive.  This iterative CISD approach developed recently, which is called adaptive configuration interaction and is unrelated to the method used in this work~\cite{Go2017}, misses some of the key features that are important for an efficient DMFT impurity solver as will be described below.

The CI methods currently used in the DMFT literature are  not representative of modern CI techniques~\cite{caffarel2016,giner2013,Robinson2017,Eriksen2017,Rahul2017,Tubman2018}. 
Selected CI (SCI) methods have recently been shown to be much more efficient than previous CI methods.   Recently, the adaptive sampling CI method (ASCI) was introduced as a modern approach to SCI, and since then the ASCI method and other approximate SCI methods have been rapidly developing further~\cite{Holmes2016,evan2016,Zimmerman2017,Zimm2017,Liu2016,Schriber2017}. The key idea that allows this method to be more efficient than traditional CI methods is to remove the active space and instead to identify the most relevant degrees of freedom in Hilbert space to describe the ground state. The ASCI method has been shown to successfully and efficiently treat strongly correlated electronic systems known for their difficulty, for example the $\mathrm{Cr}_2$ dimer~\cite{Tubman2016,Holmes2016}.

In this paper, we adapt the ASCI algorithm to act as impurity solver in zero temperature cluster DMFT calculations. In Section II we present the algorithmic details, briefly describing the DMFT loop and discussing the ASCI method and it's extension as impurity solver. In Section III we summarize the performance of the ASCI-DMFT algorithm with a detailed convergence study in the different parameters of the method and some exemplary applications to full cluster DMFT calculations. Finally, Section IV concludes with summarizing remarks and possible directions to pursue with this new tool.

\section{Methodology}

\subsection{The DMFT Formalism}

In this work, we implement the ASCI method as the impurity solver for cluster DMFT (CDMFT) simulations.
DMFT, originally proposed for studying lattice models~\cite{Metzner1989,Mueller1989a,Mueller1989b,Kotliar1992}, works by self consistently finding bath parameters for a set of sites in a sublattice~(the cluster).   
  ~Correlation effects are taken into account \cite{Kotliar1996, Kotliar2001a}, allowing for quantum fluctuations between the cluster sites and the rest of the system.  This is done by mapping the original system self-consistently into an  Anderson impurity model with Hamiltonian
\begin{eqnarray}
H_{imp} =& H_{C} + \sum_{p=1}^{N_b}\epsilon_p\  d^\dagger_{p}d_{p}\nonumber \\ &+ \sum_{p=1,\alpha=1}^{p=N_b,\alpha=N_c}\left(V_{\alpha,p}\ d^\dagger_{p}c_{\alpha} + \mathrm{h.c.}\right)\ , \label{eq:ImpurityModel}
\end{eqnarray}
 \ The parameters of this map are the number of cluster sites $N_c$, the number of bath sites $N_b$, the bath energies $\left\{\epsilon_{p}\right\}$ and the coupling terms between the bath and cluster sites $\left\{V_{\alpha,p}\right\}$. The term $H_C$ corresponds to the original system Hamiltonian restricted to the $N_c$ cluster degrees of freedom. Here and henceforth, greek indices will be used for cluster degrees of freedom, while the index $p$ will be reserved for bath degrees of freedom.

The self-consistent map begins by a choice of the bath parameters. Given the bath parameters, the ground state wave function $\ket{\psi_0}$ and energy $E_0$ of the impurity Hamiltonian are computed using the impurity solver. The impurity Green's function for the cluster degrees of freedom can then be computed \cite{Zgid2011}

\begin{eqnarray}
G_{imp}(\omega)_{\alpha,\beta} =& \bra{\psi_0}c_{\alpha}\frac{1}{\omega - (H_{imp} - E_0) + \mathrm{i}\eta}c^\dagger_\beta\ket{\psi_0} \nonumber \\ &+ \bra{\psi_0}c^\dagger_{\beta}\frac{1}{\omega + (H_{imp} - E_0) - \mathrm{i}\eta}c_\alpha\ket{\psi_0}
\label{eq:GF}
\end{eqnarray}

where $\eta$ is a small number. From that one can access the cluster self-energy 

\begin{equation}
\begin{split}
\Sigma_c(\omega)_{\alpha, \beta} =&\ \ (\omega + \mu +i\eta)\delta_{\alpha,\beta} - h_{imp,\alpha,\beta} \\ &- G_{imp}^{-1}(\omega)_{\alpha,\beta} - \Delta^{Bath}(\omega)_{\alpha,\beta},
\label{eq:SE}
\end{split}
\end{equation}

where $\mu$ is the chemical potential, $h_{imp}$ is the non-interacting part of the cluster Hamiltonian $H_C$ and we have introduced the hybridization function $\Delta^{Bath}(\omega)$. This hybridization function is the non-interaction part of the Green's function that comes from the bath degrees of freedom, and obeys the analytical expression
  
\begin{equation}
\Delta^{Bath}(\omega)_{\alpha,\beta} = \sum_{p=1}^{N_b}\frac{V_{\alpha,p}^* V_{\beta,p}}{\omega - \epsilon_p} .
\label{eq:Hyb}
\end{equation}  
  
  These impurity Green's function and self-energy are local quantities defined only on the cluster. The lattice Green's function restricted to the cluster can be computed from these local quantities according to
  
  \begin{equation}
  G(\mathbf{R}_0, i\omega) = \frac{1}{V_{BZ}}\int_{BZ}\mathrm{d}\mathbf{k}\ \left[(i\omega + \mu)-h(\mathbf{k})-\Sigma_c(i\omega)\right]^{-1},
  \label{eq:GFloc}
\end{equation}
  
  where $BZ$ denotes the first Brillouin zone and $h(\mathbf{k})$ is the momentum space representation of $h_{imp}$. The self-consistent condition then amounts to equating the Green's function of the Anderson model $G_{imp}(iw)_{\alpha,\beta}$ to the full lattice Green's function $G(\mathbf{R}_0, i\omega)_{\alpha,\beta}$. This is solved by iteratively fitting the bath parameters. The magnitude to be fit is the hybridization function $\Delta^{Calc}(\omega)_{\alpha,\beta}$, which is computed from  $G(\mathbf{R}_0, i\omega)$ in Eq.~\ref{eq:GFloc} by rewriting Eq.~\ref{eq:SE} as
  
  \begin{equation}
  \begin{split}
 \Delta^{Calc}(\omega)_{\alpha,\beta} =&\ \ (\omega + \mu +i\eta)\delta_{\alpha,\beta} - h_{imp,\alpha,\beta} \\ &- G^{-1}(\mathbf{R}_0, i\omega)_{\alpha,\beta} - \Sigma_c(\omega)_{\alpha, \beta} ,
\label{eq:HybCalc}
\end{split}
\end{equation}
  
  . This hybridization function is then fitted to the analytical expresion in Eq. \ref{eq:Hyb}, which provides with new bath parameters $\{\epsilon_p\}, \{V_{\alpha,p}\}$. In this work, we use the BOBYQA implementation in the nlopt library \cite{nlopt,bobyqa} to minimize the cost function
 
 \begin{equation}
\chi(\{\epsilon_p\},\{V_{\alpha,p}\}) = \frac{1}{N_\omega N_c^2}\sum_{n=1}^{N_\omega}\left|\Delta^{Calc}(\omega_n) - \Delta^{Bath}(\omega_n)\right| ,
\label{eq:CostFunc}
\end{equation}
 
 where we use the Frobenius norm. To fit over smooth functions, the frequencies $\omega_n$ are usually chosen along the imaginary axis. For more details and specific prescriptions on the CDMFT calculation, consult ref ~\cite{Bolech2003} and~\cite{supp}.
 
 Upon completion of the DMFT self-consistency, i.e. upon identification of the optimal bath parameters $\{\epsilon_p\}, \{V_{\alpha,p}\}$ that define the impurity Hamiltonian in Eq.~\ref{eq:ImpurityModel} encoding the low energy physics of our original system of interest, one can proceed to compute properties along the real frequency axis. In this work, we report lattice spectral weights $A(\mathbf{k},\omega)$, which can be interpreted as the momentum resolved density of states for single particle and single hole excitations. To compute this magnitude, one first has to determine the full lattice Greens function $G_{latt}(\mathbf{k},\omega)$. This is calculated by periodizing the cluster restricted Green's function. One of the usually applied periodization schemes follows
 
 \begin{equation}
G_{latt}(\mathbf{k},\omega) = \frac{1}{N_c}\sum_{\alpha,\beta=1}^{N_c}e^{i\mathbf{k}(\mathbf{r}_\alpha-\mathbf{r}_\beta)}G(\mathbf{R}_0, \omega)_{\alpha,\beta}.
\label{eq:GFlatt}
\end{equation}
 
 The spectral weights are then the imaginary part of the lattice Green's function
 
  \begin{equation}
A(\mathbf{k},\omega) = -\frac{1}{\pi}\mathrm{Im}(G_{latt}(\mathbf{k},\omega)) .
\label{eq:Akw}
\end{equation}

\subsection{The ASCI Algorithm}

To proceed, we present ASCI as an impurity solver for the cluster Hamiltonian $H_{imp}$.  There has been a lot of interest in developing CI methods to treat DMFT impurity systems, especially recently~\cite{Zgid2011,Zgid2012,Lu2014,Go2017}.   CI methods work as an impurity solver by diagonalizing a Hamiltonian in a basis of many-fermion states (determinants).  However, traditional CI methods that have been previously considered for this purpose are not the most effective to treat strongly correlated systems since they are missing several important aspects that are central to SCI.  A key feature of SCI methods is to identify the most relevant determinants needed to describe the ground state wave function. 
 In particular, ASCI does so by ranking the determinants according to their coefficient in a trial ground state wave function  and the Hamiltonian matrix elements \cite{Tubman2016, Holmes2016}. The method proceeds iteratively, improving the subspace onto which $H_{imp}$ is projected.  This subspace is referred to as the \emph{target space} and characterized by the number of determinants included, $tdets$.

ASCI starts with a guess for the target space, denoted as $\{D_{tdets}\}$, e.g., the Hartree Fock determinant plus some set of low rank excitations (singles, double, triples,...). 
The ground state energy and wave function of the Hamiltonian of interest $H$ are then computed  in the space $\{D_{tdets}\}$ (e.g., by Lanczos).  After diagonalization, the wave function is defined by its expansion coefficients $C_j$ in the $\{D_{tdets}\}$ space. The method then proceeds to update the target space by choosing a new set of determinants (a set of size $tdets$) that better describes the ground state. This update is done by searching all the singly and doubly excited determinants from a subset of $\{D_{tdets}\}$, which we denote as $\{D_{search}\}$.
  The size of $\{D_{search}\}$ is a parameter that we choose, $cdets$, and its influence on the simulation is described in detail in ref ~\cite{Tubman2016,Tubman2018}.  

 The set  $\{D_{search}\}$ contains the determinants corresponding to the largest coefficients $|C_j|$ from the ground state wave function. We denote all determinants found in the search as the set $\{D_{SD}\}$.  This set can have many orders of magnitude more elements than $\{D_{tdets}\}$.

After the search, we calculate 
\begin{equation}
A_i = \frac{\sum_{j}^{`}H_{i,j} C_j}{H_{i,i} - E_0}, \label{eq:ASCI}
\end{equation}
for all determinants in $\{D_{SD}\}$, which provides an estimate of their importance in the ground state wave function.  
The prime in the sum indicates a sum over $\{D_{tdets}\}$, $H_{i,j} = \braket{i|H|j}$, and $E_0$ is the current best estimate of the ground state energy.
 
 A new target space is built by ranking elements of the the old target space together with the new singles and doubles, according to the absolute value of their coefficients $C_j$ and $A_i$ respectively, and selecting the $tdets$ determinants with the largest coefficients. $H$ can then be generated in the new target space and its ground state computed. This process is then repeated until convergence.

This method is advantageous for systems in which the ground state can be described with enough accuracy using a small subset of the total Hilbert space. The required accuracy is application-dependent, but previous work has shown that the ASCI method can treat strongly correlated systems generally accepted as difficult, with higher accuracy and less resources~\cite{Tubman2016,Lehtola2017,Tubman2018}. The speed of convergence of the ASCI method can be greatly influenced by the correct choice of $tdets$, with larger $tdets$ yielding higher accuracy but requiring a longer time for each iteration. When dealing with a new system, one begins with a modest size of the target space and ramps this up until the ground state energy is converged to the desired precision. For a more in depth discussion of the ASCI algorithm, its other parameters and further algorithmic details to highly exploit its properties, see reference~\cite{Tubman2016,Tubman2018}.

The ASCI method provides a deterministic prescription to identify a compact representation of the ground state wave function $\ket{\psi_0}$ in a chosen basis of Slater determinants. Compact here means that ASCI identifies the most important determinants (ranked by ground state wave function coefficient) and thus reaches great energy convergence with a moderate number of determinants. 

\subsection{ASCI-DMFT}

In this paper, we adapt the ASCI algorithm to provide an impurity solver for CDMFT simulations. The main point here is that the ASCI wave function compactness translates also into a compact Green's function representation, which makes ASCI a time and memory efficient CI-based impurity solver.

For this we need to perform an additional step in order to calculate the Green's function in Eqn. \ref{eq:GF} efficiently. The Hamiltonian $H_{imp}$ has to be inverted once in the basis with one particle more than in $\ket{\psi_0}$ and  once in the basis with one particle less. In a CI approach, these bases for the Hilbert spaces of single particle and hole excitations on top of the ground state target space $\{D_{tdets}\}$ have to be truncated. Naively, one could construct these spaces simply by applying the corresponding creation (annihilation) operators on the converged $\{D_{tdets}\}$ basis, in order to be able to represent single particle (hole) excitations. We denote these naive bases as $\{D^\pm_{tdets}\}$. These bases would be enough to represent the impurity Hamiltonian in the single particle (hole) spaces to the accuracy of the ASCI wave function. However, to compute the impurity Green's function we need to invert the Hamiltonian. To represent the inverse of the Hamiltonian to the accuracy of the wave function, we need more states because by inverting we shuffle all matrix elements communicating with the target space $\{D_{tdets}\}$. The solution to this problem is to add the states connected to the single particle (hole) space by the Hamiltonian, which for any quartic Hamiltonian corresponds to adding single and double excitations on top of the naive bases. The coefficients of these states in the expansion of the corresponding single particle (hole) states are zero in our level of approximation, so we call them zero states. Adding these zero states forms the final target space for the Green's function calculation $\{D^{+,Z}_{tdets}\}$.

For clarification, we give a simple example. Let us consider a Hilbert space where each state is characterized by 5 fermionic modes. In second quantized notation, each state is then characterized by the 5 occupation numbers, and can be written as $\ket{n_1n_2n_3n_4n_5}$. We will assume that the target space $\{D_{tdets}\}$ of the ASCI calculation in this system has 2 fermions. A possible state would be $\ket{00011}\in \{D_{tdets}\}$. When computing the Green's function matrix element $(\alpha,\beta) = (1,1)$ in Eqn. \ref{eq:GF} we will act with $c^\dagger_1$. The corresponding element of $\{D^+_{tdets}\}$ to $\ket{00011}$ is $\ket{10011}\in \{D^+_{tdets}\}$. To compute the inverse of $H_{imp}$ more accurately, we now complete $\{D^+_{tdets}\}$ by adding singly and doubly connected states, the zero states. Now, which states are singly or doubly connected to  $\{D^+_{tdets}\}$ depends on $H_{imp}$. For the sake of this example, let us assume that the Hamiltonian only includes single excitations to neighboring sites in a 1D line, which is what would happen in a 1d Hubbard chain. Then, for state $\ket{10011}$ we would only need to add $\ket{01011}$ and $\ket{10101}$. After all the pertinent inclusions, we have formed the final basis $\{D^{+,Z}_{tdets}\}$.

Along the imaginary frequency axis, looking for one set of zero states is usually enough to converge the Green's function, at least for the systems presented in the next section. Along the real frequency axis, to converge the pole structure of the Green's function one has to add more than one set of zero states. In particular for the calculations presented in the Results section, we needed to include all zero states connected to the $\{D^{+,Z}_{tdets}\}$ set described above. For the Hubbard model, that only includes single excitations in the spatial basis, this means that we added the single excitations of the original naive space $\{D^+_{tdets}\}$ and then added single excitations of those single excitations. In this work, we refer to this as adding two "layers" of zero state excitations. With increasing correlation in the Hamiltonian, it is expected that further layers of exciations will be needed to get converged Green's functions.

This method becomes more costly with increasing number of degrees of freedom, i.e. with increasing $N_c$ and $N_b$ in the DMFT method. The scaling is essentially exponential. To avoid prohibitively large $\{D^{+,Z}_{tdets}\}$ sets, we perform a truncation in the same spirit as done in previous configuration interaction impurity solvers \cite{Go2017}. The main idea is to only include zero states connected to the leading $m$ determinants in the ASCI ground state wave function, ordered by the absolute value of the coefficient. This critically reduces the size of the $\{D^{+,Z}_{tdets}\}$ spaces and allowed to add up to $N_c + N_b = 40$ spinful degrees of freedom (i.e. containing $2\cdot (N_c + N_b)$ spin orbitals), the most complicated system being a $N_c = (4\mathrm{x}4)$ with $N_b = 24$. We have found that adding all states with an absolute ground state coefficient larger than $10^{-4}$ is enough to converge all Green's functions presented in this work, while keeping the size of $\{D^{+,Z}_{tdets}\}$ always bellow 10 million states. In the next section, we refer to the size of $\{D^{+,Z}_{tdets}\}$ as $GFtdets$.

\section{Results}

To demonstrate the efficiency of the ASCI method as an impurity solver in CDMFT calculations, we consider here the one-dimensional (1d) and two-dimensional (2d) square lattice Hubbard models. The Hubbard model is characterized by the Hamiltonian

\begin{eqnarray}
\label{eq:Hubbard} H_{Hub} = &-t \sum_{\langle i,j \rangle, \sigma}\left( c^{\dagger}_{i,\sigma}\ c_{j,\sigma} + h.c. \right)  -\mu \sum_i (n_{i,\uparrow} + n_{i,\downarrow}) \nonumber \\ &+  U\sum_{i} n_{i,\uparrow}\ n_{i,\downarrow}, 
\end{eqnarray}

\noindent with hopping amplitude $t$, chemical potential $\mu$, Coulomb interaction strength $U$ and spin label $\sigma = \{\uparrow, \downarrow\}$. This Hamiltonian, reduced to a number of sites, is what enters as $H_C$ in Eqn. \ref{eq:ImpurityModel}. At half filling we have $\mu = U / 2$, for other particle fillings one would need to determine the chemical potential and the number of electrons in the impurity model self consistently according to \cite{Zgid2011}.

By optimizing the target space in the ASCI method, we show that we can reproduce the results in the literature with drastically reduced computational resources. We first show the convergence behavior of the ASCI impurity solver for 1D and 2D Hubbard models as a function of the target space size $tdets$ and the total basis size for the $\{D^{\pm,Z}_{tdets}\}$ spaces.

\subsection{Convergence Tests}

\subsubsection{Hamiltonian Truncation}
\label{sec:res_conv}

\begin{figure}
\includegraphics[width=0.5\textwidth]{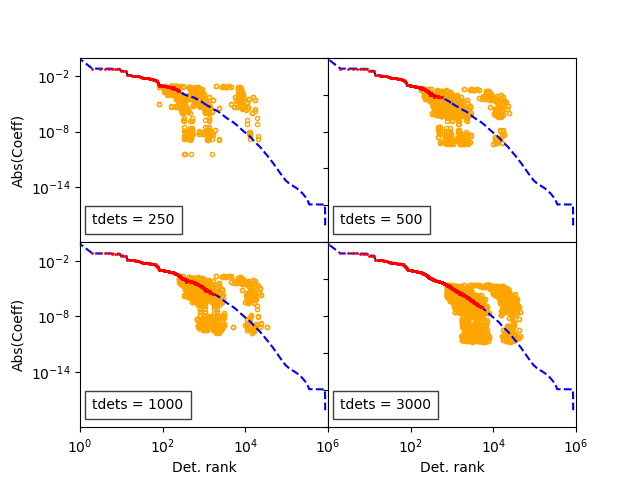}
\caption{\label{fig:1Dcoeffs} Sorted absolute values of the ground state wavefunction coefficients for the final iteration in a 1D Hubbard model DMFT calculation with $U / t = 8$, $N_c = 1$, $N_b = 11$. The determinant order is determined by the exact diagonalisation calculation, represented in all sub-figures by the blue dashed line. Each sub-figure presents the corresponding wavefunction coefficients for ASCI-DMFT calculations using different $tdets$, namely $tdets = $ 250, 500, 1000 and 3000 as red dots. The estimated coefficients as computed according to Eq. \ref{eq:ASCI} are presented as orange circles.}
\end{figure}

\begin{figure}
\includegraphics[width=0.5\textwidth]{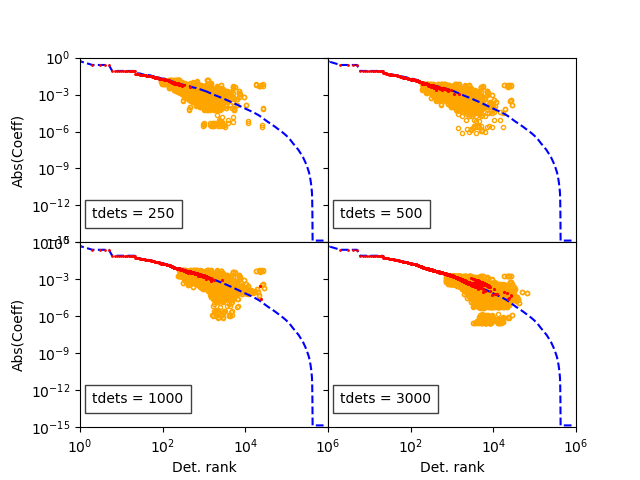}
\caption{\label{fig:2Dcoeffs} Sorted absolute values of the ground state wavefunction coefficients for the final iteration in a 2D Hubbard model cluster DMFT calculation with $U / t = 8$, $N_c = 2\mathrm{x}2$, $N_b = 8$. The determinant order is determined by the exact diagonalisation calculation, represented in all sub-figures by the blue dashed line. Each sub-figure presents the corresponding wavefunction coefficients for ASCI-DMFT calculations using different $tdets$, namely $tdets = $ 250, 500, 1000 and 3000 as red dots. The estimated coefficients as computed according to Eq. \ref{eq:ASCI} are presented as orange circles.}
\end{figure}

We first show that ASCI can indeed identify the most important determinants to describe the ground state wave function for the typical impurity models that are encountered in DMFT calculations. For that, we present the sorted absolute values of the ground state wave function coefficients for two systems that can be solved with exact diagonalisation (ED), both at half filling. i) A 1D Hubbard model DMFT calculation with $N_c=1$ and $N_b=11$, Fig. \ref{fig:1Dcoeffs}. ii) A 2D square lattice Hubbard model cluster DMFT calculation with $N_c=2\mathrm{x}2$ and $N_b=8$, Fig. \ref{fig:2Dcoeffs}.  The ED coefficients are shown as blue dashed lines, and the coefficients computed in ASCI calculations with different $tdets$, namely $tdets = $ 250, 500, 1000 and 3000, are shown as red dots. As a reference, the total number of states in the full Hilbert space is 853776. Additionally, we report the estimated coefficients $A_i$ for the search set $\{D_{SD}\}$ in Eq. \ref{eq:ASCI} of the main text as orange circles. In both figures, the ability of ASCI to select the most important $tdets$ determinants becomes completely evident, and the coefficients computed with ASCI (red dots) show excellent agreement with the ED results (blue dashed line). Both of these overlay the estimation coefficients (orange circles), which become less accurate the further away from the currently explored region of the Hilbert space. The only minor discrepancies arise for the states with smallest coefficients when the target space includes more 3000 states for the 2D system. In this case, ASCI seems to have more difficulties to adapt to the abrupt decrease in the coefficients from the $10^4$-th state onwards. These difficulties arise probably from the higher degree of strong correlation in this two dimensional, cluster calculation. The estimated $A_i$ coefficients, shown as orange circles in the Figures, have a greater discrepancy with the ED results which in some cases can be of some orders of magnitude. However, these estimates follow the general shape of the ED coefficients well and allow for the efficient and accurate identification of the most relevant states. ASCI is shown thus to be able to select the most important states to describe the ground state for the kind of impurity models that arise in DMFT calculations.
  
\begin{figure}
\includegraphics[width=0.5\textwidth]{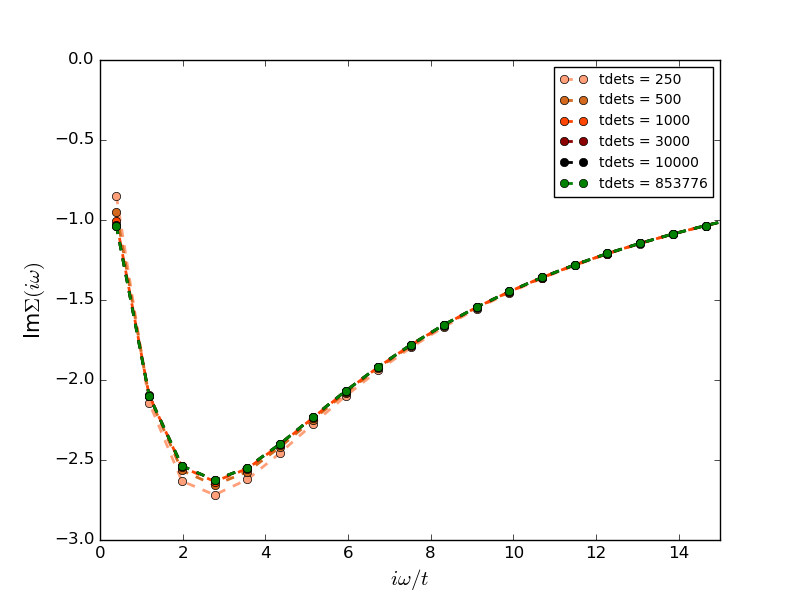}
\caption{\label{fig:1D_Nc1} Imaginary part of the self-energy along the imaginary frequency axis $\mathrm{Im}\Sigma(i\omega)$ for a 1d Hubbard impurity model at half-filling. ASCI results with $U / t = 8$, $N_c = 1$, $N_b = 11$ and different sizes of the target space. Presented are $tdets = $ 250, 500, 1000, 3000 and 10000 in different scales of red, and the exact diagonalisation results in green.}
\end{figure}

\begin{figure}
\includegraphics[width=0.5\textwidth]{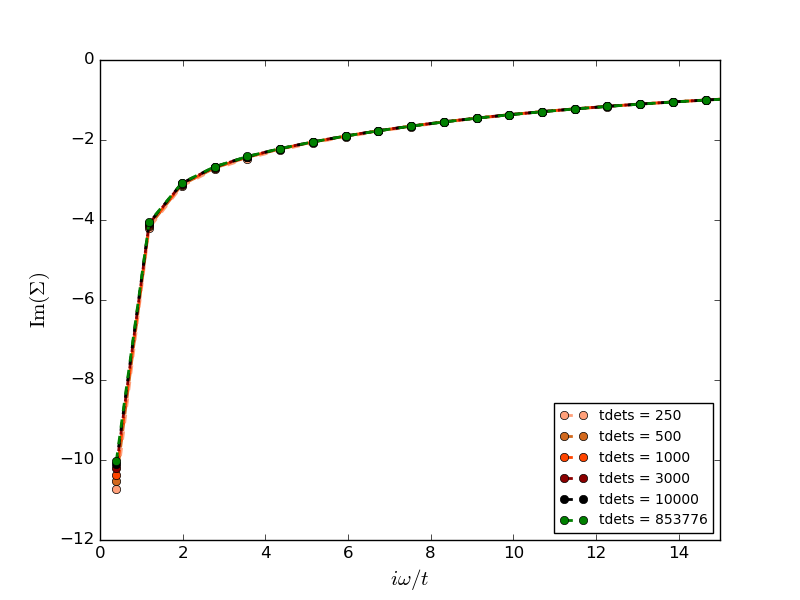}
\caption{\label{fig:2D_Nc2x2} Imaginary part of the diagonal self-energy along the imaginary frequency axis $\mathrm{Im}\Sigma(i\omega)_{0,0}$ for a 2d Hubbard impurity model at half-filling. ASCI results with $U / t = 8$, $N_c = 2\mathrm{x}2$, $N_b = 8$ and different sizes of the target space. Presented are $tdets = $ 250, 500, 1000, 3000 and 10000 in different scales of red, and the exact diagonalisation results in green.}
\end{figure}

\begin{figure}
\includegraphics[width=0.5\textwidth]{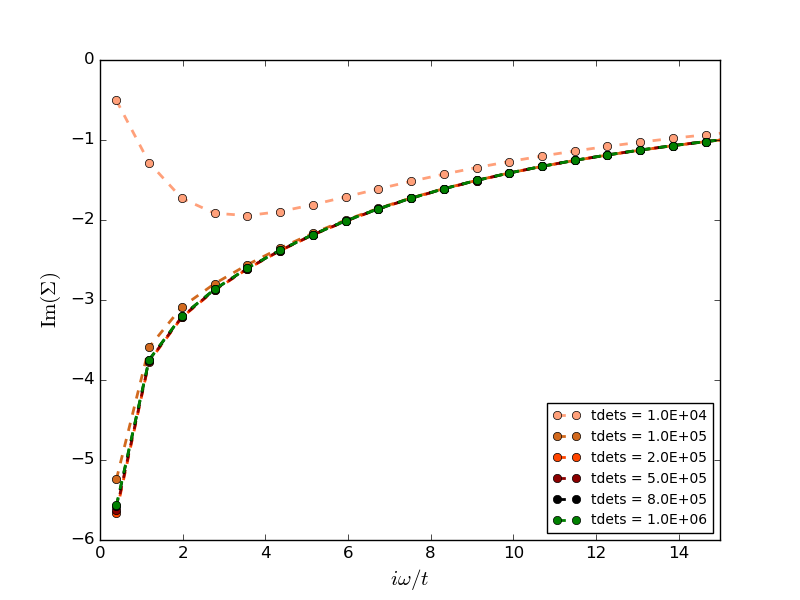}
\caption{\label{fig:2D_Nc3x3} Imaginary part of the diagonal self-energy along the imaginary frequency axis $\mathrm{Im}\Sigma(i\omega)_{0,0}$ for a 2d Hubbard impurity model at half-filling. ASCI results with $U / t = 8$, $N_c = 3\mathrm{x}3$, $N_b = 19$ and different sizes of the target space. Presented are $tdets = $ 10000, 100000, 200000, 500000 and 800000 in different scales of red, and one million determinants in green.}
\end{figure}

\begin{figure}
\includegraphics[width=0.5\textwidth]{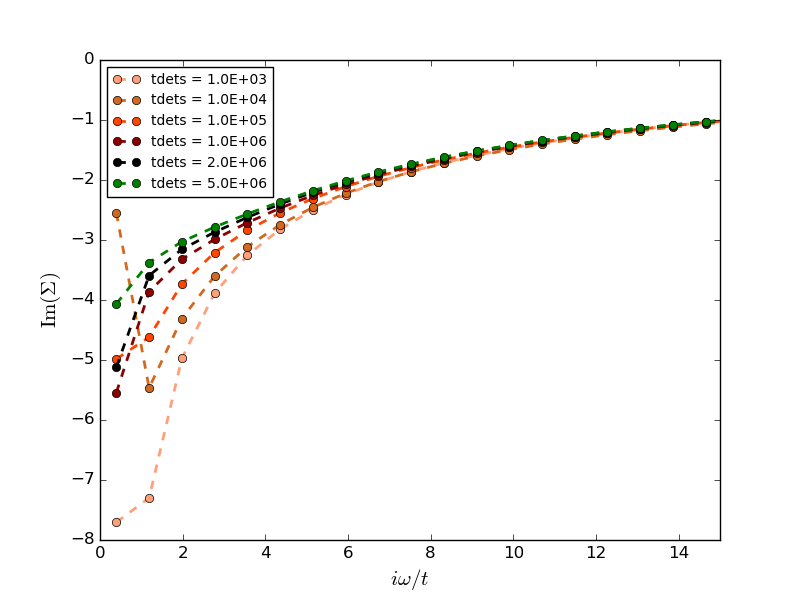}
\caption{\label{fig:2D_Nc4x4} Imaginary part of the diagonal self-energy along the imaginary frequency axis $\mathrm{Im}\Sigma(i\omega)_{0,0}$ for a 2d Hubbard impurity model at half-filling. ASCI results with $U / t = 8$, $N_c = 4\mathrm{x}4$, $N_b = 24$ and different sizes of the target space. Presented are $tdets = $ 1000, 10000, 100000, 1000000, 2000000 in different scales of red, and five million determinants in green.}
\end{figure}

The ASCI method can thus provide a compact and accurate ground state wave function representation, converging the wave function coefficients fairly rapidly. To assess it's proficiency as an impurity solver, it is pertinent to assess how frequency dependent functions like the self energy converge with the number of determinants $tdets$. We present the convergence of the first diagonal element of the self energy for different impurity problems: a 1D Hubbard model with $N_c = 1$ and $N_b = 11$ in Fig. \ref{fig:1D_Nc1}, a 2D $N_c = 2\mathrm{x}2$ and $N_b = 8$ in Fig. \ref{fig:2D_Nc2x2}, a 2D $N_c = 3\mathrm{x}3$ and $N_b = 19$ in Fig. \ref{fig:2D_Nc3x3} and a 2D $N_c = 4\mathrm{x}4$ and $N_b = 24$ in Fig. \ref{fig:2D_Nc4x4}, all at half-filling and $U / t = 8$. Where possible, we perform ED calculations for comparison. When necessary, we truncate the Green's function bases $\{D^{\pm,Z}_{tdets}\}$ to a maximum of ten million states. The effect of the truncation of the $\{D^{\pm,Z}_{tdets}\}$ spaces is presented in the following subsection.

We observe a very rapid convergence of the self energy with $tdets$ in almost all cases. Indeed, the order of $10^5$ determinants seem to be enough to converge the qualitative behavior in the low frequency regime. A significant quantitative difference at low frequency between the different calculations with $tdets \leq 10^5$ is only appreciable in the largest impurity cluster, $N_c = 4\mathrm{x}4$ in Fig. \ref{fig:2D_Nc4x4}, which are not converged with the target space sizes used in this work. The success with the $3\mathrm{x}3$ cluster suggests, however, that further improvement of the algorithm will make convergence in this challenging cluster possible. In particular, we want to draw attention to the fact that the current implementation of the ASCI impurity solver is not exploiting any active space structure, which has been noted to be fundamental for configuration interaction based solvers~\cite{Go2015,Lin2013,Go2017}. Including this kind of structure will further boost the convergence, by reducing the effective number of orbitals to the active space, which results in an exponential reduction of the Hilbert space size ASCI searches through. 

\begin{figure}
\includegraphics[width=0.5\textwidth]{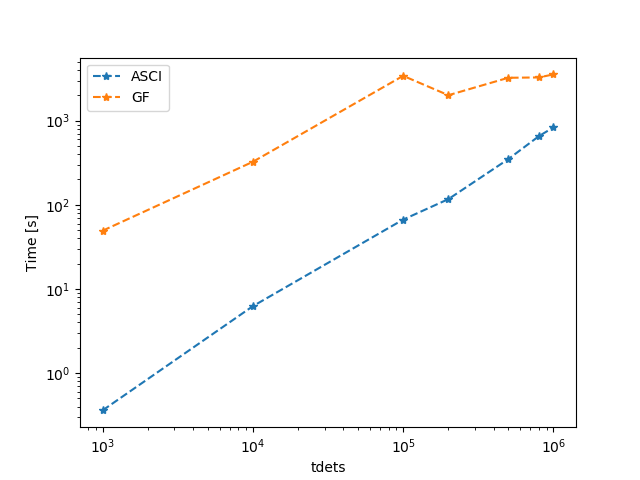}
\caption{\label{fig:timings_3x3} Time in seconds for the ASCI ground state calculation (blue) and the Green's function computation (orange) for the $N_c =3\mathrm{x}3$, $N_b = 19$ 2d Hubbard impurity model at half-filling and $U / t = 8$ as a function of the target space size $tdets$.}
\end{figure}

The timings for the different parts of the computation deserve consideration. Fig. \ref{fig:timings_3x3} shows the timings for the ASCI procedure and for the Green's function calculation for the $N_c = 3\mathrm{x}3$ impurity model calculations. Converging the ground state energy and wave function can be done under 15 minutes for all the impurity models presented here, while computing the Green's function elements can take up to 1 hour per element in the largest systems. These timings are orders of magnitude better than those reported for equivalent CI based zero temperature DMFT solvers \cite{Go2017}. In fact, the bottle-neck in the ASCI-DMFT procedure is now the fitting step for the large cluster calculations, as reported below. To increase the range of applicability of the ASCI-DMFT algorithm it is thus imperative to improve upon the fit methodologies.

\subsubsection{Green's Function Truncation}

\begin{figure}
\includegraphics[width=0.5\textwidth]{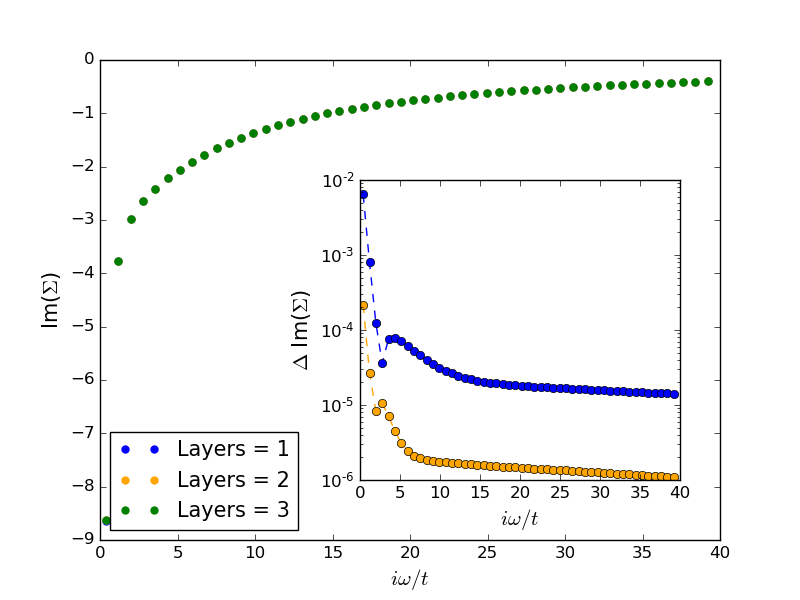}
\caption{\label{fig:GF_trunc} Imaginary part of the diagonal self energy along the imaginary frequency axis $\mathrm{Im}\Sigma(i\omega)_{0,0}$ for a 2d Hubbard impurity model at half-filling. ASCI results with $U / t = 8$, $N_c = 2\mathrm{x}2$, $N_b = 24$ with $tdets = 5\cdot 10^5$ for different truncation schemes of the Green's function spaces $\{D^{\pm,Z}_{tdets}\}$. The curves correspond to adding singles to the naive Green's function space once, twice and up to three times. See Methods section for details. Inset shows the absolute difference between the results with one layer (blue) and two layers (orange) with the three layer calculation.} 
\end{figure}

In order to ascertain convergence in the truncation of the $\{D^{\pm,Z}_{tdets}\}$ spaces, we report the self energy for $tdets~=~5\cdot10^5$ and different truncation schemes in the intermediate size 2D cluster $N_c = 2\mathrm{x}2$ and $N_b = 24$ in Fig.~\ref{fig:GF_trunc}. We report the truncation as the number of layers of zero states included in the $\{D^{\pm,Z}_{tdets}\}$ space, performing calculations with one, two and three layers. See Methods section for details.

As shown in Fig. \ref{fig:GF_trunc}, the convergence behavior is rapid along the imaginary frequency axis with the number of layers, the results being quantitatively converged already with a single layer. Adding the second layer, which corresponds to two orders of magnitude more zero states in the $\{D^{\pm,Z}_{tdets}\}$ space, does not change the self energy in any significant way. This property is inherited from the compact ASCI ground state wave function.

\begin{figure}
\includegraphics[width=0.5\textwidth]{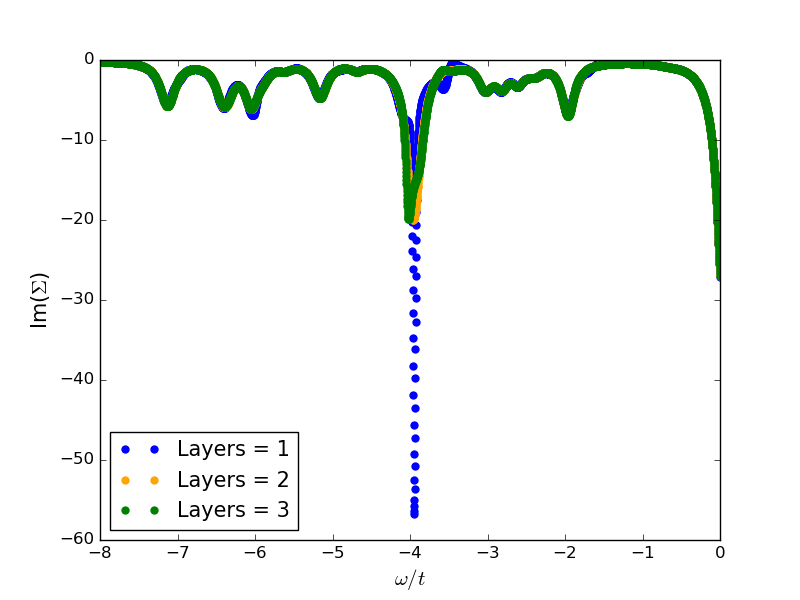}
\caption{\label{fig:GFtdets_effect} Imaginary part of the diagonal self energy along the real frequency axis $\mathrm{Im}\Sigma(\omega)_{0,0}$ for a 2d Hubbard impurity model at half-filling. ASCI results with $U / t = 8$, $N_c = 2\mathrm{x}2$, $N_b = 24$ with $tdets = 5\cdot 10^5$ for different truncation schemes of the Green's function spaces $\{D^{\pm,Z}_{tdets}\}$. The curves correspond to adding singles to the naive Green's function space once, twice and up to three times. See Methods section for details.} 
\end{figure}

All the convergence results presented above concern the calculation along the imaginary frequency axis, where the DMFT loop takes place. As described in the methods section, upon conclusion of this loop one finds the impurity Hamiltonian that best describes the low energy physics of the original lattice model. To extract these physical properties however, one has to perform one final calculation along the real frequency axis. Although this obviously does not change the convergence requirements for the ground state target space $\{D_{tdets}\}$, the description of the poles of the Green's function can and in fact does increase the necessary size of the truncated space $\{D^{\pm,Z}_{tdets}\}$. As an illustrative example, we present the spectral weights for a $N_c= (2\mathrm{x}2)$, $N_b = 24$ calculation at half-filling with $U/t = 8$ and three different truncation schemes, corresponding to adding different layers of zero state excitations, in Fig. \ref{fig:GFtdets_effect}.

The example of Fig. \ref{fig:GFtdets_effect} illustrates the claims made in the methods section: the pole structure of the Green's function makes the convergence along the real frequency axis more demanding, so that we need to include at least two layers of zero state excitations to converge all peaks. It is important to note however that this is a small number of layers compared to equivalent impurity solvers based on selective configuration interaction methods~\cite{Go2017}. Moreover, in the case presented in Fig. \ref{fig:GFtdets_effect} all self energies are causal, even the unconverged ones computed with the minimal number of zero state layers. This is in strong contrast with the method in~\cite{Go2017}, where it is necessary to go up to 4 layers of zero state excitations to achieve causality in impurities of comparable size. This difference comes from the more efficient identification of relevant states for the ground state in the ASCI solver, which in turn translates in needing only a minimal amount of states for the Green's function representation. Since the number of additional zero states scales exponentially with the number of layers, this improvement is very relevant to allow access to larger impurity and bath sizes. This is especially so considering that larger impurities may require a larger number of layers.

For all the calculations in the following section, we used one layer along the imaginary frequency axis and two layers along the real frequency axis.

\subsection{cDMFT Results}


Having established the timing and convergence properties of the ASCI algorithm as an impurity solver for the Hamiltonians that naturally arise from cluster DMFT calculations in the 1D and 2D Hubbard models, we now show example applications of the ASCI-DMFT to study the two-dimensional square lattice Hubbard model. Here, we choose $U / t = 8$ at half filling. When away from half-filling, one needs to undertake a self-consistent determination of the chemical potential and number of electrons that (a) minimize the energy, and (b) represent the desired lattice filling \cite{Zgid2011}. Using current CI based DMFT methods for this self-consistent calculation is excessively expensive in time, and most benchmarking has therefore been done at half-filling. Given the timings and scaling presented above, the ASCI impurity solver can also be used to speed up those kind of calculations. We limit our presentation to half-filling for reasons of brevity. 

\begin{figure*}
\begin{center}
\scalebox{1}
{\includegraphics[width=1.\textwidth]{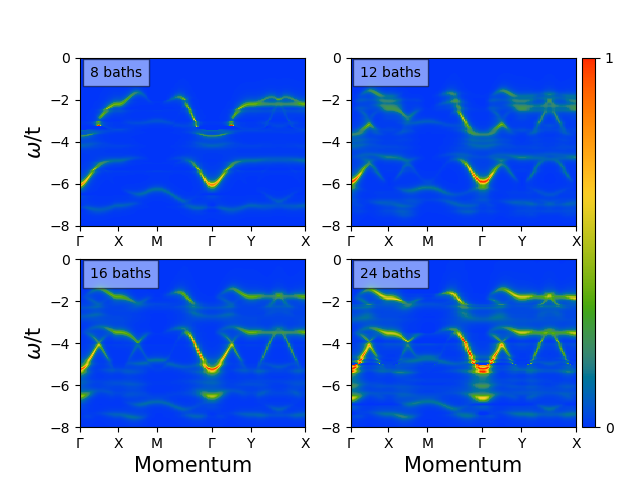}}
\end{center}
\caption{\label{fig:akw_2x2_baths} Spectral weights $A(\mathbf{k},\omega)$ for the two-dimensional square lattice Hubbard model in a calculation with $U / t = 8$, $N_c = (2\mathrm{x}2)$ and $N_b = 8, 12, 16$ and 24 respectively. We show a particular cut through the first Brillouin zone. The abrupt change observed for the 16 and 24 bath calculations is due to instabilities inherent to the fitting process of the hybridization function, see Eq. \ref{eq:CostFunc}. See main text for details.}
\end{figure*}

In the case of the small 2x2 cluster, the compact wave function representation of the ASCI impurity solvers allows us to study the effect of the bath discretization error by performing simulations with $N_b =$8, 12, 16 and 24 bath sites. The spectral weights for these simulations are presented in Figs. \ref{fig:akw_2x2_baths}. The $N_b=8$ calculations can be done in 1 hour on a single core and show excellent agreement with previous literature \cite{Go2017}, while the $N_b=12,\ 16$ and 24 calculations required 15, 48 and 65 hours respectively. These timings include the complete DMFT calculations, which is performed in the imaginary frequency axis, but do not account for the final computation of the Green's function along the real frequency axis. Due to the multiple singularities along the real frequency axis, the Hamiltonian inversion required in Eqn. \ref{eq:GF} is extremely numerically demanding and required the use of a parallelized Lanczos routine to compute all Green's function elements in under 24 hours.

 We see that the spectral weights can change noticeably with the number of baths. In general, all calculations share the same main features as the $N_b = 8$ case (upper left panel in Fig. \ref{fig:akw_2x2_baths}). The calculations with larger baths seem to include more light bands than the $N_b = 8$ case, diminishing very slightly the particle weight of the main bands. In particular, there is an inverted parabola between the Y and X points that becomes more and more pronounced with a larger bath. Beside that, there is little change in the shape of the main features, which present an almost quadratic dispersion in the vicinity of the $\Gamma$ point.

There is however a drastic change in the spectral weights when going from 12 baths to 16 baths. The lower hole band shifts to smaller energies by almost a full energy unit $t$, see lower pannels in Fig. \ref{fig:akw_2x2_baths}. While this change does not affect the insulating gap appreciably, it is nevertheless unexpected that the convergence in the number of bath sites would show such a step change midway. This slow convergence behavior with increasing number of baths is due to the inherent instability of the fitting procedure with increasing $N_b$ along the imaginary frequency axis, and not to an issue with the impurity solver. Concretely, the large bath solution with shifted lower bands stems from over-fitting the long frequency behavior of the hybridization function in Eq. \ref{eq:Hyb} at the cost of an accurate description of its low frequency behavior, where the particular physics are encoded. As a consequence, increasing the number of baths is making the DMFT self-consistency iteration converge to a different fix point. While it is standard in the literature to introduce a cutoff in the fit, and only consider the very small frequency behavior, e.g. \cite{Foley2018}, it would be more desirable to use a fitting method capable to account for both the small and large frequency domains at the same time. The authors are currently working on a collaboration to devise and characterize an efficient and robust fitting method, and have observed that while $N_b = 16$ is too small a bath to account for the full frequency range, it is possible with $N_b = 24$ when using an appropriate fit \cite{Mejuto2019}. 

The impact of such a fitting method goes beyond just allowing the study of the large bath limit in small clusters. When treating impurity clusters with many degrees of freedom, the number of bath parameters to be determined by the fit increases correspondingly, making the fitting process the more complicated and unstable if done without care. Using complex bath couplings in the Hamiltonian in Eqn. \ref{eq:ImpurityModel}, the number of real fitting parameters grows as $2N_cN_b$, which for the 3x3 cluster with 17~baths corresponds to 306 real fitting parameters to fit a 9x9 complex, frequency dependent matrix. This is a very demanding task for a fitting procedure, and devising a robust and scalable method for this is far from trivial. In this work, we used the BOBYQA implementation in the nlopt library \cite{bobyqa,nlopt}. The fitting procedure becomes the bottleneck of our calculation for the 3x3 clusters and for a 4x4 DMFT loop it requires an impracticable amount of time, needing on occasions up to 24 hours to perform one fit. In these circumstances, the quality of the fit has to be put under severe scrutiny and a search for more reliably fitting procedures becomes imperative. 
 
\begin{figure}
\includegraphics[width=0.5\textwidth]{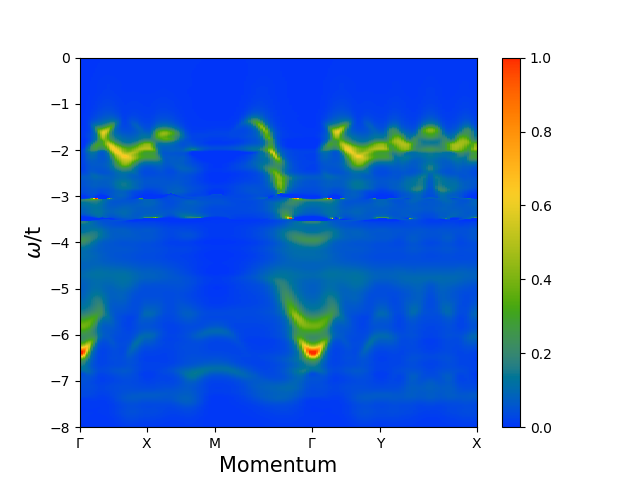}
\caption{\label{fig:akw_3x3} Spectral weights $A(\mathbf{k},\omega)$ for the two-dimensional square lattice Hubbard model in a calculation with $U / t = 8$, $N_c = (3\mathrm{x}3)$ and $N_b = 17$. We show a particular cut through the first Brillouin zone.}
\end{figure}

Using ASCI as an impurity solver allows us to make first calculations with cluster sizes larger than the current state of the art for CI based DMFT methods. Fig.~\ref{fig:akw_3x3} presents spectral weights for the 3x3 cluster and $N_b=17$ at half-filling. This calculation took approximately twenty hours. The computational bottleneck  as mentioned above is the fitting step for the bath parameters. Increasing the number of cluster and bath sites dramatically increases the number of fitting parameters, which makes the non-linear fitting process expensive, an issue that the authors are currently addressing \cite{Mejuto2019}. 

The results for 3x3 with $N_b=17$ agree well with the 2x2 results. The system shows the insulating behavior and main features seen for the 2x2 $N_b=8 $ calculations. However, consistent with the larger bath size as noted above, the spectral weights show a plethora of small side features, just as in the 2x2 $N_b=12, 16$ and 24 calculations. It is important to note that the spectral weights in Fig. \ref{fig:akw_3x3} are not quantitatively converged with respect to the Green function target space size, due to limitations in the final real axis calculation. This can be seen in the small dots of high intensity at $\omega\approx 3\ t$. However, this convergence issue should not affect the position of the features, mainly the sharpness of the bands. Additionally, these effects are relatively far away from the Fermi level, which is centered around $\omega = 0$ in our figures. Thus, we are confident of the qualitative picture shown in Fig. \ref{fig:akw_3x3}. Further improvements on the ASCI algorithm, in particular regarding the implementation of active space reduction as already discussed in Sec.~\ref{sec:res_conv} will smooth out these small mismatches.

\section{Conclusion and Outlook}

We have presented a CDMFT implementation using ASCI as the impurity solver, and shown that 
the superior efficiency of this approach allows study of both the bath discretization error in small clusters and exploration of cluster sizes beyond the current state of the art for CI based DMFT methods. The results provide strong motivation for undertaking further application of modern CI techniques to DMFT.  Additional tools available with ASCI include many-body perturbation theory corrections~\cite{Tubman2018a}, heat-bath extensions~\cite{Holmes2016} and the exploitation of active space structures~\cite{Lin2013}, all of which can be expected to accelerate these algorithms for applications of DMFT. A new suite of algorithms for increasing the efficiency of ASCI on modern computers will improve the timings presented here even further~\cite{Tubman2018}. 
ASCI can also be readily applied to other embedding techniques such as density matrix embedding theory~\cite{Knizia2012,Zgid2015,Medvedeva2017}.

We demonstrated the effect of the main parameter of the ASCI method in the truncation approach, the size $tdets$ of the active space. The physical properties of the system along the imaginary frequency axis, represented by the cluster self energy, converge very quickly with a modest number of determinants. The convergence on the real frequency axis requires a larger space, but is also fairly rapid. A reasonable strategy is thus to use small to moderate target space sizes for the DMFT iteration loops, which occur along the imaginary frequency axis, and then to increase the size on the real axis for computing the measurable physical properties of the system. Parameters may be further tuned during the iterations along the imaginary frequency axis by beginning with a small number of determinants, computing a few DMFT loops to bring the bath parameters into the correct range, and then increasing $tdets$ to achieve high accuracy in the fits. This is particularly important for calculations away from half-filling, in which the number of electrons and the chemical potential have to be determined in a self-consistent procedure. Thus, at the beginning of the self-consistent method one could start with a small number of determinants, which can then be increased once the desired particle filling is reached. At the large bath or cluster limit, the development of efficient and robust fitting methods is still necessary \cite{Mejuto2019}. 

Application of ASCI to molecular Hamiltonians, has allowed simulation on the order of 50 electrons in 100-200 sites/orbitals~\cite{Tubman2018}.  For DMFT applications, we aim to further develop the ASCI impurity solver to further increase the number of degrees of freedom that it can handle. 
Our main future goal is the study of complicated systems, e.g., many-band Hubbard models for the study of transitions between exotic phases of matter \cite{Clay1999} or realistic many body Hamiltonians, for example by combining our solver with \emph{ab initio} methods such as GW-EDMFT \cite{Nilsson2017}.
We envision this new efficient impurity solver having the potential to also be useful for benchmarking the solution of embedding Hamiltonians with hybrid quantum-classical algorithms realized on quantum computers~\cite{Bauer2016,Rubin2016}.

\section*{Acknowledgements}
We acknowledge helpful discussions with James LeBlanc and Olivier Parcollet.   CMZ thanks the Fundaci\'on Bancaria La Caixa for a \emph{Obra Social ``La Caixa''} graduate fellowship.
NMT  was  supported  through  the  Scientific  Discovery  through  Advanced  Computing  (SciDAC)  program
funded  by  the  U.S.  Department  of  Energy under  Contract
No.  DE-AC02-05CH11231.  Computational  resources  provided  by  the  Extreme  Science  and  Engineering  Discovery
Environment (XSEDE), which is supported by the National
Science Foundation Grant No. OCI-1053575, are gratefully
acknowledged.

%

\pagebreak
\clearpage

\onecolumngrid

\section*{Supporting Information}

\setcounter{figure}{0}  
\setcounter{equation}{0}

Here we present bullet point flow charts for the DMFT and ASCI algorithms. ''Ground state'' is shortened as GS. A brief physical motivation of the impurity model is also in order: In DMFT, quantum fluctuations outside from the cluster sites is accounted for by introducing the fermionic bath site degrees of freedom. The possibility for a particle to leave the cluster, move through the rest of the lattice and finally return to the cluster is thus represented by the couplings to the single bath sites. From this interpretation it becomes clear that an infinite number of bath degrees of freedom is formally needed to recover the thermodynamic limit behavior of the system. Since exact diagonalization or configuration interaction solver force finite baths, studying the effect of this bath discretization error becomes fundamental to evaluate the validity of DMFT calculations.

\begin{table}[h!]
\caption{DMFT method \cite{Zgid2011}\label{tab:DMFT}}
\begin{ruledtabular}
\begin{tabular}{ll}
  \underline{Input}:& Number of cluster sites $N_c$, Number of bath sites $N_b$, Hamiltonian $H$. \\
  \underline{Output}:& Bath parameters $E_p$, $V_{p,\alpha}$ ($p \in [1, N_b], \alpha \in [1,N_c]$).\\
  \underline{Algorithm}:& \\
  & 1. Initial guess for bath parameters. Usually this comes from a low level calculation, like Hartree-Fock. \\
  & 2. Compute GS wave function $|GS\rangle$ and energy $E_{GS}$ of current impurity model with the impurity solver. \\
  & 3. Compute cluster Green's function $G_c(i\omega)$ and self energy $\Sigma_c(i\omega)$. These are $(N_c\ \mathrm{x}\ N_c)$ matrices defined as \\
  &\ \ \ \ $G_{c,(\alpha,\beta)}(i\omega) = \langle GS| c_\alpha \frac{1}{i\omega + \mu - (H - E_{GS})} c_\beta^{\dagger}|GS\rangle + \langle GS| c_\beta^\dagger \frac{1}{i\omega + \mu + (H - E_{GS})} c_\alpha|GS\rangle$ and $\Sigma_c(i\omega) = G_0^{-1}(i\omega) - G_c(i\omega)$,\\
  &\ \ \ \ where $c_\alpha$, $c_\beta^\dagger$ are the annihilation and creation operators for cluster sites $\alpha$ and $\beta$ respectively, $G_0(i\omega)$\\
  &\ \ \ \ is the non-interacting Green's function of the impurity model and $\mu$ is the chemical potential.\\
  & 4. Compute local full lattice Green's function from the cluster Green's function by Fourier transforming\\
  &\ \ \ \ $G(i\omega,\mathbf{R}) = \frac{1}{V_{BZ}}\int_{BZ}\mathrm{d}\mathbf{k}\ \exp\left(i\mathbf{k}\cdot\mathbf{R}\right)\left[(i\omega + \mu)-h(\mathbf{k})-\Sigma_c(i\omega)\right]^{-1}$. Here $h(\mathbf{k})$ is the Fourier transform of the\\
  &\ \ \ \ non-interacting part of $H$ with respect to the unit cell defined by the cluster and $BZ$ stands for the Brillouin \\
  &\ \ \ \ zone defined by the same unit cell. To recover the Green's function on the cluster, we set $\mathbf{R}=\mathbf{R_0}\equiv \mathbf{0}$.\\
  & 5. Impose self-consistency $G_c(i\omega) = G(i\omega,\mathbf{R_0})$. This means that we can express $G(i\omega,\mathbf{R_0})$ with the bath parameters.\\
  & 6. Find new bath parameters by fitting $G(i\omega,\mathrm{R_0})$. \\
  & 7. If bath parameters converged, finish. If not, go to step 2. \\
\end{tabular}
\end{ruledtabular}
\end{table}

\begin{table}[h!]
\caption{ASCI method \cite{Tubman2016,Tubman2018} \label{tab:ASCI}}
\begin{ruledtabular}
\begin{tabular}{ll}
  \underline{Input}:& Size $tdets$ of the target space $\{D_{tdets}\}$, size $cdets$ of the core space $\{D_{search}\}$, Hamiltonian $H$, basis (usually\\
  
  &localized orbital basis). \\
  \underline{Output}:& Optimal target space of size $tdets$, ground state energy $E_{GS}$ and wavefunction $|GS\rangle$.\\
  \underline{Algorithm}:& \\
  & 1. Initial guess for target space. This can be a Hartree-Fock solution plus single and double excitations. \\
  & 2. Compute GS in current target space. \\
  & 3. Find connected singles and doubles to the $cdets$ most important target space states (the core space).  Most \\
  &\ \ \ \ important means largest coefficient in the GS wavefunction. \\
  & 4. Rank all states, target space plus the singles and doubles from the core space, according to Eq. \ref{eq:ASCI}. \\
  & 5. Update the target space by choosing the top $tdets$ states of the ranking in step 4. \\
  & 6. Compute GS in the new target space. \\
  & 7. If GS energy is converged, finish. If not, go to step 3. \\
\end{tabular}
\end{ruledtabular}
\end{table}

\begin{table}[h!]
\caption{This work - ASCI-DMFT (Calculating the Green's function) \label{tab:ASCI-DMFT}}
\begin{ruledtabular}
\begin{tabular}{ll}
  \underline{After ASCI}: & Find spaces for particle and hole excitations. This is needed to compute the cluster Green's function as \\
  & shown in Table \ref{tab:DMFT}. We need to find the zero states connected to each of the $c_\alpha^\dagger\{D_{tdets}\}$ and $c_\alpha\{D_{tdets}\}$, where\\
  & $\alpha$ runs over all cluster sites. For the diagonal elements of $G_c(i\omega)$, one proceeds as:\\
  & 1. Act with the creation operator on the ASCI target space. The states reached this way form the \\
  &\ \ \ \ core of the single particle  excitation space.\\
  & 2. Find all zero states connected to the $m$ leading states and add them to the target space. Zero states are\\
  &\ \ \ \  single and double excitations on top of the states found in step 8. Rather than fixing a number $m$, we \\
  &\ \ \ \ introduce a lower cutoff for the absolute value of the ground state coefficient. We only find zero states for \\
  &\ \ \ \ states with absolute coefficients above the cutoff. Additional layers of zero are added until convergence is \\
  &\ \ \ \ reached. An additional layer of zero states means to add the zero states connected to the zero states of a \\
  &\ \ \ \ previous layer. For the systems studied in this work, one layer is enough along the imaginary frequency axis, \\
  &\ \ \ \ while the real axis requires two layers.\\
  & 3. Proceed analogously with the annihilation operator for the single hole excitation \\
  & For the off-diagonal element $(\alpha,\beta)$ of $G_c(i\omega)$, one proceeds by computing the Green's function for $c^\dagger_\alpha\pm c^\dagger_\beta$.\\
  & Additional states might need to be added if there are zero states connected to the $(\alpha,\alpha)$ space that can only be \\
  &reached from the $(\beta,\beta)$ space and vice-versa.\\
 & Adding these steps to the ASCI algorithm presented in Table \ref{tab:ASCI} provides a functional impurity solver for DMFT. \\
\underline{Example:}& Consider a 1 dimensional system of spinless fermions with\\
&five sites for the fermions to reside. Any state in this system of fermions can be described in second quantization as\\
&$|n_1n_2n_3n_4n_5\rangle$, where $n_i$ is the occupation number for site $i$. The number of particles $N= \sum_{i=1}^5n_i$ is an integer \\
&between 0 and 5. Let us assume that the target space $\{D_{tdets}\}$ of the ASCI calculation has 2 fermions. A possible \\
&state would be $|00011\rangle \in \{D_{tdets}\}$. When computing the Green's function matrix element $(1,1)$, we will act with $c_1^\dagger$\\
&on $\{D_{tdets}\}$. This will map $|00011\rangle \to |10011\rangle$. $|10011\rangle$ is an element of the core single particle excitation space for\\
&the  (1,1) element of the Green's function (In contrast, when computing the corresponding single hole excitation space, \\
&the state $|00011\rangle$ would not contribute, since it is mapped to zero by the annihilation operator $c_1$). Now, to complete\\
&the single particle excitation space, we want to add zero states, i.e. N+1 (in this case 3) particle states that are\\
&connected by the Hamiltonian to  $c_1^\dagger\{D_{tdets}\}$. Those are states with $n_1 = 0$. Now, which states are connected to the  \\
&core single particle excitation space depends on the nature of the Hamiltonian. For the sake of our example, we will \\
&assume that $H$ only includes nearest neighbor hopping terms, connecting state $i$ with states $i+1$ and $i-1$ (with \\
& periodic boundary conditions). Thus, from $|10011\rangle$ we can only reach the zero states $|01011\rangle$ and $|10101\rangle$, any other  \\
&3-fermion state with $n_1=0$ cannot be reached from $|10011\rangle$ by nearest neighbor hops. Thus, we would only add \\
&$|01011\rangle$ and $|10101\rangle$.This process has then to be repeated for each state in $\{D_{tdets}\}$ and for each creation and anni- \\
&hilation operator.\\

\end{tabular}
\end{ruledtabular}
\end{table}

\end{document}